\documentstyle[prb,aps,epsf,twocolumn]{revtex}
\newcommand{\BE}{\begin{equation}}
\newcommand{\EE}{\end{equation}}
\newcommand{\BA}{\begin{eqnarray}}
\newcommand{\EA}{\end{eqnarray}}

\begin{document}
\twocolumn[\hsize\textwidth\columnwidth\hsize\csname@twocolumnfalse\endcsname
\title{Effect of dielectric responses on localization in 1D random periodic-on-average systems}
\author{\small Cheng-Ching Wang \thanks{Email: ccwang@phys.cts.nthu.edu.tw}\\
{\footnotesize \it Physics Division, National Center for
Theoretical Sciences, {\it P. O. BOX} 2-131, Hsinchu, Taiwan
30043}\\ \small Pi-Gang Luan \thanks{Email:
pgluan@cc.nctu.edu.tw}\\ {\footnotesize \it Institute of
Electrophysics, National Chiao-Tung University, Hsinchu, Taiwan
30043}}
\date{\small December 28, 2001}

\maketitle

\begin{abstract}

Dielectric response effects on wave localization in random
periodic-on-average layered systems (POAS) are studied. Based on
Monte Carlo simulations and products of Random Matrices,
statistics of the Lyapunov exponent are determined efficiently for
very long systems. A novel oscillatory behavior for Lyapunov
exponent is found and explained for mildly strong scattering
conditions. We also show the emergence of strongly localized
states in metallic layered systems with intermediate disorder for
frequencies above the plasma frequency $\omega_{p}$ of metals, as
is not shown in dielectrics. Furthermore, the violation of
universal single parameter scaling behaviors in different regimes
of multiple scattering is discussed.\vspace{3mm}
\end{abstract}]

The realization that Anderson localization in electronic
systems\cite{Anderson2} is due to wave interference between
scattered waves from random scatters  has stimulated vivid
research in search of wave localization in condensed matter
physics\cite{Ping,Laijk,Nori,Sigalas,Zhen}. Further progress have
been made to the aspects of photon
localization\cite{John,Joan,E.Y,Mara1,Vlasov,Torre,AA,Deych1,Deych2,Deych3,Gang}
because photons may open a new realm of optical and microwave
phenomena, and provided an analogous insight into Anderson
localization transition undisturbed by Coulomb interaction.

It had been shown individual scatter's response to the wave fields
influence wave localization properties remarkably in entirely
random systems , such as the strong localization for acoustic
waves in bubbly liquids\cite{Zhen}. The same issues should exist
in POAS for the intricate interplay between order and disorder.
One-dimensional systems are particular interesting as they provide
insights to the problems of wave localization in general and are
suitable for testing various ideas. Two qualitatively different
regimes of localization are exhibited\cite{Vlasov,Deych1}. For
band gap states, single parameter scaling (SPS) with universal
behaviors is observed; however, the scaling is restored only when
the randomness of defects exceeds a certain threshold for the
situation of weak dielectric mismatch between constituent
layers\cite{Deych1}.

The scope of the Communication is twofold. The main interest is to
understand multiple scattering effects on wave localization and
the SPS in 1D systems. Particular attentions are paid for mildly
scattering conditions; that is, the dielectric mismatch between
layers is not too large. The secondary interest lies in
preliminary exploration of dispersive or absorptive media on wave
localization. Here, we consider EM wave localization in 1D random
superlattices made of two alternating layers, the so-called
Kronig-Penny model. As usual, the wave transmission can be tackled
in exact manner by the transfer-matrix method\cite{Deych1}.
Nevertheless, the approach encounter the difficulty that
transmission coefficient in that formulation falls bellow the
computer round-off accuracy for very long systems\cite{Deych3}.
For the practical purpose, we improve the Herbert-Jones-Thouless
formula\cite{book1} widely known in the study of disordered
electronic quantum systems to study electromagnetic wave
localization, with electron energies being replaced by wave
frequencies; consequently, universal behaviors for very long
structures can be studied quantitatively. Numerically, the state
of wave fields is formulated by a vector
$v_{n}=(E_{n},E^{'}_{n})^t$ with the electric field $E_n$, its
spatial derivative $E^{'}_{n}$, the wave number $k=\omega/c$ in
vacuum, and the superscript $t$ represents the matrix
transposition. The relation between the (n+1)th layer and the nth
layer is specified by the transfer matrix ${\rm M_{n}}$:
\begin{equation}\label{eq:2}
{\rm M}_{n}=\left(
 \begin{array}{clr}
 {\rm cos}(k_{n}d_{n})& \frac{1}{k_{n}}{\rm sin}(k_{n}d_{n}) \\
 -k_{n}{\rm sin}(k_{n}d_{n}) & {\rm cos}(k_{n}d_{n})
 \end{array}
 \right)
\end{equation}
where $k_{n}=k\sqrt{\epsilon_{n}(\omega)}$, and the n-th layer
width $d_{n}$ . In dielectrics, the dielectric constant
$\epsilon_{n}$ is real; however, in metals, the dielectric
constant $\epsilon_{n}$ is complex and frequency-dependent.
 Lyapunov exponent(LE) and its variance (VAR),
$\mbox{Var}(\lambda)=\langle\lambda^{2}\rangle-\langle\lambda\rangle^{2}$,
are the transmission quantities considered here. The VAR
represents the fluctuations for the LE in defect configurations.
The LE can be computed as follows \cite{book1}:
\begin{equation}
\label{eq:3} \lambda = {\rm Re} \lim_{N \rightarrow \infty}
\frac{1}{L} \left\langle \ln \left( {\rm Tr} \prod_{n=1}^{2N}{\rm
M}_{n}({\rm Conj}(\epsilon_{n}))
 \right)\right\rangle
\end{equation}
where $2N$ is the total number of alternating layers, $L$ is the
average total length of considered systems, Tr is the trace
operator, Conj means taking complex conjugate, and the symbol
$\langle ... \rangle$ means the ensemble average over the uniform
distribution of random configurations. The complex conjugate of
the dispersive dielectric function $\epsilon_{n}$ is taken for the
reciprocity of the absorptive systems \cite{Deych3}.

Consider 1D superlattices made of two alternating layers A and B
with the dielectric response functions $\epsilon_A$ and
$\epsilon_B$ respectively. Random configurations are introduced by
varying the layer $B$ width $d_{B}$ uniformly in the interval
$(\langle d_{B}\rangle (1-\sigma),\langle d_{B} \rangle
(1+\sigma))$ , where $\sigma\in(0,1)$ represents the randomness
while the width of $A$ layer $d_{A}$ is kept fixed. A series of
Random Matrices ${\rm M}_{n}$ are generated to obtain the LE and
VAR by the Monte Carlo simulation in Eq.(2). In practice, the
accurate LE and VAR are attained by letting the total length of
layered systems to be several thousand times larger than the
decaying length-scale $\lambda^{-1}$ (the inverse of Lyapunov
exponent), representing exactly the localization length in
absorption free cases. Sufficient times of ensemble averages are
thus determined by the evaluation of numerical stability with the
length of the system.
\input epsf.tex
\begin{figure}
\epsfxsize=3.2in \centerline{\epsffile{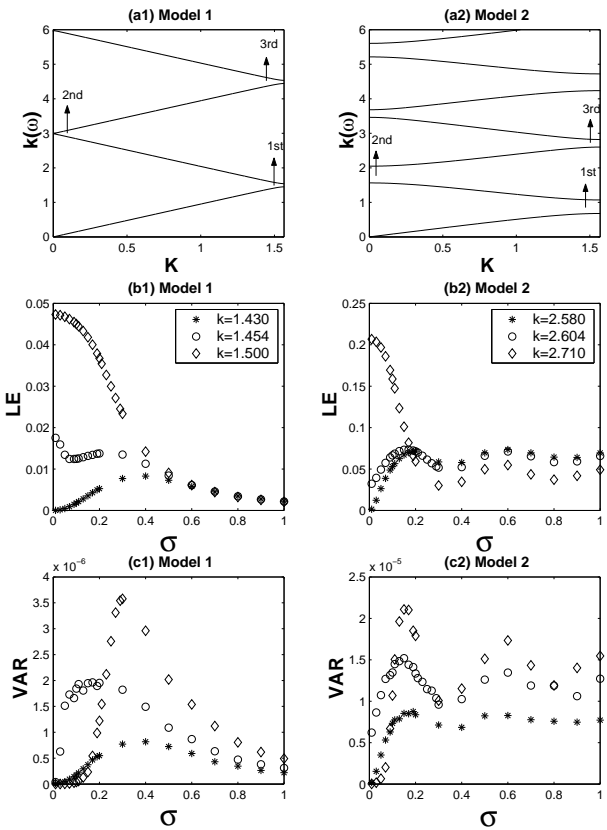}} \caption{Model
1: (a1) Band structures where K is Bloch wave number in periodic
layers and ${\rm k} = \omega/c$ is the wave number in free space.
(b1) Lyapunov exponent vs. randomness $\sigma$. (c1) Variance vs.
randomness $\sigma$. Model 2 : (a2) Band Structures. (b2) Lyapunov
exponent vs. randomness $\sigma$. (c2) Variance vs. $\sigma$.}
\end{figure}
To begin with, multiple scattering effects on wave localization in
dielectrics are studied. Two models are demonstrated. Model 1 is a
weakly scattering model with the dielectric contrast
$\epsilon_{B}/\epsilon_{A}=1.21$ while Model 2 is the mildly
strong scattering model with the contrast
$\epsilon_{B}/\epsilon_{A}=6.25$. The band structures of original
periodic layers\cite{Bloch} for Model 1 and 2 are shown in Figs. 1
(a1) and (a2). The LE and VAR in Model 1 and 2 are compared in
Figs. 1(b1 - c1) and (b2 - c2). The parameters, $\epsilon_{A}=1,
d_{A} = \langle d_{B}\rangle=d=1$, are taken in computations. The
symbols `$*$',`$\circ$', and `$\diamond$' denote states in bands,
band edges, and gaps respectively hereafter.

In Model 1, previous results are reproduced\cite{Deych1}. In Fig.
1 (b1), the LE decreases monotonically with randomness in the
first gap around $k=1.50$ since the established Bragg wave
coherence has been gradually destroyed, the so-called enhanced
transmission due to disorder\cite{Mara1}. In the allowed band
$k=1.43$, the LE initially increases with the randomness and then
decreases for large randomness. In the band edge $k=1.454$, the LE
decreases first, then increase slightly, and finally decrease
monotonically with the randomness $\sigma$. The behavior is due to
the interplay between order and disorder in the systems. Moreover,
the curves for these states merge for large disorder, the
manifestation of universal single parameter scaling\cite{Deych1}.
In Fig. 1 (c1), the plot of VAR vs. randomness $\sigma$ is
demonstrated. The VAR at these frequencies shows similar
behaviors. When the randomness of defects is large in the
ensembles, the effect of self-averaging is effective because
random fluctuations for these states are almost cancelled leading
to the reduction in VAR. For nearly periodic configurations, the
wave localization length in a gap regime is much smaller than the
mean distance between defects so that the defects are isolated for
incident waves. The increase of randomness can only cause minor
growth of fluctuations due to the increase of the localization
length. The competition between the two mechanisms results in the
observed maximal VAR.
\begin{figure}
\epsfxsize=3.2in \centerline{\epsffile{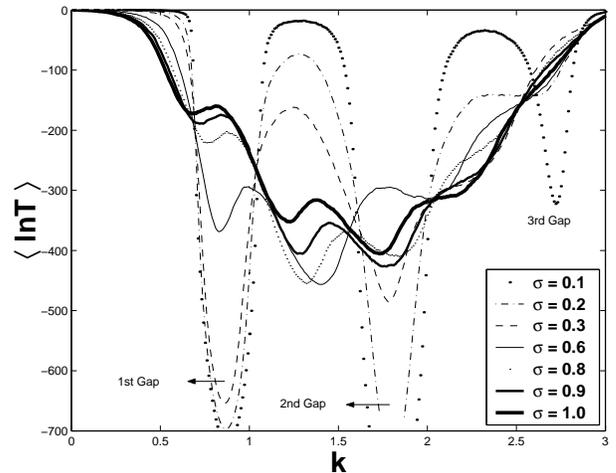}} \caption{The
average transmission $\langle {\rm lnT}\rangle = -2{\lambda}N$ vs.
frequency ${\rm k}$ with different randomness $\sigma$.}
\end{figure}
In Model 2, mildly strong scattering case, novel oscillatory
behaviors of the LE and VAR in the 3rd gap are exhibited above the
randomness $\sigma=0.3$ in Figs. 1 (b2) and (c2), which is very
different from the behaviors of Model 1 in Figs. 1 (b1) and (c1).
To understand the disorder effects, we plot the average
transmission spectrum ${\rm \langle lnT \rangle}$ covering the
lowest three pass bands and gaps of Model 2 for the randomness
$\sigma$ from $0.1$ to $1$, as is shown in Fig. 2. First, we
observe well defined band structures, transmission enhancement in
gaps, and inhibition in pass bands appears as usual when the minor
randomness ranging from $\sigma=0.1$ to $\sigma=0.3$ is presented.
With large randomness above $\sigma=0.3$, we observe the enormous
overlap between original band structures emerges from the high
frequency states from $k=2$ to $3$ and then extends to the states
from $k=1$ to $3$. In addition, it is noticed that the oscillatory
behavior only occurs for these states. Take the 2nd gap for
example, the transmission initially enhance till the randomness
$\sigma=0.6$, and then weaken for larger randomness. For the 3rd
gap, the oscillation in the transmission corresponds to the
oscillation in LE in Fig. 1 (b2). For the first gap and pass band
where the overlap with other states is negligible, the behavior
does not exist. Therefore, the novel oscillation in LE could be
explained as a result of the overlap between band structures in
these states for large randomness.
\begin{figure}
\epsfxsize=3.2in \centerline{\epsffile{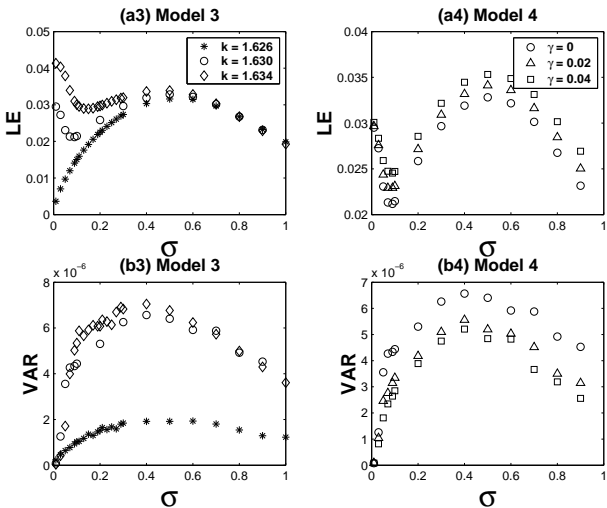}} \caption{Model
3: (a3) Lyapunov exponent vs. randomness $\sigma$. (b3) Variance
vs. randomness $\sigma$ . Model 4: (a4) Lyapunov exponent vs.
randomness $\sigma$ at the frequency $k=1.630$ in Model 3. The
circle $\rm o$, triangle $\triangle$, and square $\Box$ represent
the cases with electron damping rates $\gamma$ = $0$, $0.02$, and
$0.04$, respectively. (b4) Variance vs. randomness $\sigma$.}
\end{figure}
Turn to the secondary interest. As a preliminary inspection on
dispersive media, the statistics of LE and VAR for metallic slabs
are demonstrated in Fig. 3. The layers $B$ are composed of metals
with the dielectric response function $\epsilon_B:$
 \begin{equation}\label{eq:1}
\epsilon_B(\omega)
=1-\frac{{\omega_{p}}^{2}}{\omega(\omega+i\gamma)}
\end{equation}
where $\omega_{p}$ is the plasma frequency, $\omega$ the EM wave
frequency, and $\gamma$ the electron damping rate which leads to
residual wave energy dissipation. The fixed parameters for layer
$A$ are the same as previous Model 1 and 2. The only difference is
the introduction of the plasma frequency $\omega_{p}d/c=1$ ( $c$ :
speed of light in vacuum) of metallic slabs, and the reduced wave
number $k=\omega d/c=\omega/\omega_{p}$.
\begin{figure}
\epsfxsize=3.2in \centerline{\epsffile{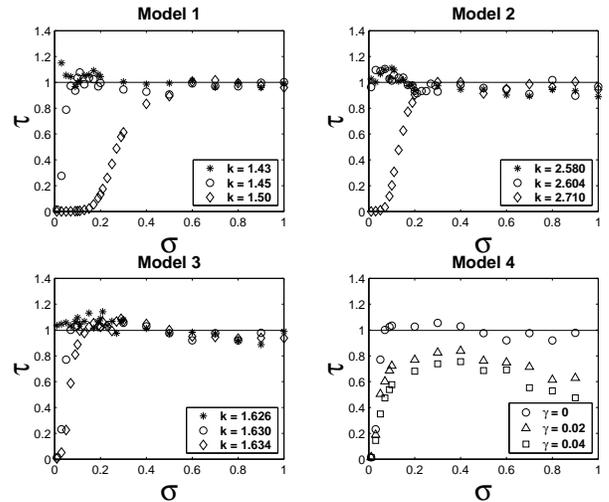}} \caption{Scaling
parameter $\tau$ vs. randomness for Model 1, Model 2, Model 3, and
Model 4, respectively. The horizontal solid line denotes the
universal value $\tau = 1$.}
\end{figure}
We focus on frequencies above the plasma frequency $\omega_{p}$
since waves can only propagate above the plasma frequency. It is
noted that the dielectric contrast $\epsilon_{B}/\epsilon_{A}$ in
this case is real in the absence of residual absorption. The
residual absorption can be turned off by taking electron damping
rate $\gamma$ as zero. The frequencies in later computations are
taken from the first gap above $\omega_{p}$. In Fig. 3 (a3), a
peculiar feature not shown in dielectrics occurs. For the gap
state $k=1.634$, the contrast between the LE at the randomness
$\sigma \simeq 0$ and 0.5 is much smaller than Model 1 (See Fig. 1
(b1)). Moreover, the LE is even larger with the intermediate
randomness $\sigma = 0.5$ than the LE with the minor randomness
$\sigma\simeq 0$ at the edge state $k=1.630$. The explanations are
as follows. The dispersive response of metals makes the dielectric
contrast between layers lower for higher frequencies, which leads
to the reduction of extra transmission difference between the pass
bands and gaps. So, the feature is attained by diminishing the
difference of these states in LE in its dielectric counterpart at
minor randomness $\sigma \simeq 0$ in qualitative agreement with
the behavior in Fig. 3 (a3). The effect is remarkable with nearly
periodic layers because individual layer's responses can be
accumulated by the Bragg wave coherence. This suggests versatile
ways could be available in the fabrication of the EM wave devices
operating near band edges in POAS with dispersions\cite{John1}. In
Fig. 3 (b3), the VAR agrees qualitatively with Model 1 although
the shape of the VAR is not so sharp as Model 1. In addition, the
VAR does not merge for these frequencies at large degrees of
disorder, the implication of the SPS violation to be discussed
later. One may wonder the peculiar feature may be smeared out with
residual absorption. To confirm the point, the LE vs. randomness
figure is plotted in Fig. 3 (a4) as comparison. We see the feature
can be strengthened further because the LE increases noticeably
with the damping rate $\gamma$ for large randomness. The VAR
decreases with the damping rate in Fig. 3 (b4), as is shown
similarly in the reference\cite{Deych3} .

Single parameter scaling (SPS) is studied next to understand its
dependence on constituent scatter properties, as is shown in Fig.
4. The SPS indicates that the LE is not an independent parameter
and is related to its variance in a universal way
\cite{Anderson2,Deych2}:
 \begin{equation}
 \tau = \frac{\mbox{Var}(\lambda)}{\lambda}L = 1,
\end{equation} where $L$ is the length of the system.
The horizontal solid line represents the universal value $\tau =
1$. In Fig. 4, for the weakly scattering case, the SPS is obeyed
for all band gap states in Model 1. This is reasonable since weak
gaps make the macroscopic wave coherence easily broken by
disorders, as is shown by Deych et al\cite{Deych2}. However, the
SPS is not rigorously met for large randomness with slight
fluctuations in $\tau$ in Model 2 and Model 3. The phenomenon is
universal in the two models, one is dielectric and the other is
dispersive, even though their detailed LE and VAR disagree. In
retrospection, in Model 3, the dispersive dielectric response in
metals generates an intrinsically larger dielectric mismatch than
Model 1. Therefore, the degree of deviations near the universal
value $\tau=1$ can be recognized as the signature of the gradual
dominance of multiple scattering effects. In Model 4, we show that
the parameter $\tau$ decreases with electron damping rate as well
as randomness. Moreover, the residual absorption makes the value
of the parameter ${\tau}$ smaller than one. From above
discussions, the SPS depends much on varied dielectric responses
of constituents. In weak scattering cases, the SPS is met for
allowed bands, gaps, and edges. In mildly scattering cases, the
parameter shows deviated fluctuations near the universal value
$\tau = 1$. However, in residual absorption cases, the parameter
$\tau$ falls below the universal value.

In addition, for strongly scattering conditions, the SPS is
violated for two possibilities in our investigations. One is due
to narrow band structures created in underlying periodic layers
resulting in tremendous overlap between these bands in the
presence of randomness\cite{Gang}. The other possibility checked
by us, however, is the generation of wide bands by exchanging the
sequence of the same constituent layers such that the overlap does
not emerge and gaps are robust enough to resist the addition of
randomness. Therefore, to sum up, the band structures of
underlying periodic layers in various scattering regimes
determines the distinctive wave localization properties in POAS.

In conclusion, we have studied effects of dielectric response on
wave localization and single parameter scaling behaviors in
periodic-on-average systems. Novel oscillatory wave localization
behaviors in the presence of mildly strong multiple scattering
effects is found and explained numerically. For dispersive
dielectric response, it is also shown that a strong localization
occurs for intermediate degree of disorder in metallic layers,
which was not found previously in dielectrics. The findings
suggest the possible applications over optical devices operating
near photonic band edges in dispersive periodic-on-average
structures in parallel with photonic crystals\cite{John1}.
Furthermore, the SPS is not rigorously met in the presence of
multiple scattering effects or residual absorption in our studies.

The work was supported by National Center for Theoretical Sciences
(Physics Division) in Hsinchu, Taiwan.


\begin{thebibliography}{99}
\bibitem{Bloch}
$\cos KD=\cos k_{A}d_{A}\cos k_{B}d_{B} - \cosh 2\eta \sin
k_{A}d_{A}\sin k_{B}d_{B}$ where $D=d_{A}+d_{B}$, $K$ is the Bloch
wave number , $k_{A}$ is the wave number in the layer A, $k_{B}$
is the wave number in the layer B, and
 $\cosh 2\eta \equiv
\frac{1}{2} \left( \sqrt \frac{\epsilon_{B}}{\epsilon_{A}}+\sqrt
\frac{\epsilon_{A}}{\epsilon_{B}} \right)$.
\end{thebibliography}
\end{document}